\documentstyle[aps]{revtex}
\begin{document}
\tightenlines
\draft
\title{Primordial Fluctuations in the
Warm inflation scenario with a more realistic coarse - grained field}
\author{Mauricio Bellini\footnote{E-mail address: mbellini@mdp.edu.ar}}
\address{Departamento de F\'{\i}sica, Facultad de Ciencias Exactas  y
Naturales \\ Universidad Nacional de Mar del Plata, \\
Funes 3350, (7600) Mar del Plata, Buenos Aires, Argentina.}
\maketitle
\begin{abstract}
I study a semiclassical approach to warm inflation scenario introduced
in previous works. In this work, I define the fluctuations for
the matter field by means of a new coarse - grained
field with a suppression factor $G$. This field
describes the matter fluctuations on the now observable scale of
the universe.
The power spectrum for the fluctuations of the matter field is
analyzed in both, de Sitter and power - law expansions for the universe. 
The constraint for the spectral index gives a constraint for the mass
of the matter field in the de Sitter expansion and a constraint for the
friction parameter in the power - law expansion for the universe.
\end{abstract}
\vskip 2cm
\noindent
PACS number(s): 98.80.Cq, 05.40.+j, 98.70.Vc
\vskip 2cm

\section{Introduction}
The inflationary universe scenario asserts that, at some very
early time, the universe went through a de Sitter phase
expansion with scale factor $a(t)$ growing as $a \sim e^{H_o \  t}$. 
Inflation
is needed because it solves the horizon, flatness, and 
monopole problems of the very early universe and also provides 
a mechanism for the creation of primordial density fluctuations\cite{pri}.
Quantum fluctuations\cite{1} 
and thermal fluctuations\cite{BF}
of matter fields play a prominent role in inflationary cosmology. 
They lead to density perturbations that would 
be responsible
for the origin of structures in the universe\cite{lin}. 
Structure formation scenarios, in particular, can receive important
restrictions based on the measured $\delta T_r /<T_r> = 1.1 \  10^{-5}$.
According to the
standard inflation
model, the formation of large - scale structure in the universe has
its origin in the growth of primordial inhomogeneities in the matter 
distribution.

The standard slow - roll inflation model separates expansion
and reheating into two distinguished time periods. It is first
assumed that exponential expansion from imflation places the universe
in a supercooled phase. Subsequently thereafter the universe
is reheated. Two outcomes
arise from such a secenario. First, the required density
perturbations in this cold universe are left to be
created by the quantum fluctuations of the inflaton. Second,
the temperature cliff after expansion requires a temporally
localized mechanism that rapidly distributes sufficient
vacuum energy for reheating.

The warm inflation scenario takes into account
separately, the matter and radiation energy fluctuations. 
In this scenario the fluctuations of the matter field 
lead to perturbations 
for matter and radiation
energy densities\cite{nphys,prl98},
which are responsible for the fluctuations of temperature.
The field $\varphi$ interacts with the
particles which are in a thermal bath with a mean temperature
smaller than the GUT critical temperature, $<T_r> \  < T_{GUT} \sim 10^{15}$
GeV. This scenario was introduced by A. Berera\cite{be1}, and developed
in other work\cite{be2}. 
In the Berera's formalism the
temperature of the universe is constant during inflation and
the temporal evolution for the amplitude of the fluctuations of temperature
are not considered. 

In the warm inflation
era, the kinetic component of
energy density $\rho_{kinetic}$ must 
be small with respect to the vacuum energy,
which is given by the potential $V(\varphi)$
\begin{displaymath}
\rho(\varphi) \sim \rho_m \sim V(\varphi) \gg \rho_{kinetic}.
\end{displaymath}
where
\begin{displaymath}
\rho_{kinetic} = \rho_r(\varphi)+ \frac{1}{2} \dot\varphi^2,
\end{displaymath}
and the radiation energy density is
\begin{displaymath}
\rho_r(\varphi) = \frac{\tau(\varphi)}{8H(\varphi)} \dot\varphi^2.
\end{displaymath}
Here, $\varphi$ is a scalar field of matter, 
$\tau(\varphi)$ is a friction parameter due to the interaction of
the matter field $\varphi$ with other fields of the thermal
bath with temperature $T_r$. Furthermore, the dot denotes the
derivative with respect to the time.
The conventional treatment of scalar field dynamics assumes
that it is pure vacuum energy dominated. The various kinematic
outcomes are a result of specially
chosen Lagrangians. In most cases the Lagrangian is unmotivated
from particle phenomenology. Clear exceptions are the 
Coleman - Weinberg potential with a untuned coupling
constant, which is motivated by grand unified theories and
supersymmetric potentials\cite{cw}. Making an
extension to the warm inflation picture, the
behavior of the scale factor can also be altered for
any given potential when the radiation energy is present.

In an alternative formalism of warm inflation\cite{plb98} I demonstrated
that, for a power - law inflation model, the
amplitude for both, thermal fluctuations and mean temperature, 
decrease
with time for a sufficiently rapid
expansion of the scale factor of the universe.
Thus, at the end of inflation the spectrum of the 
coarse - grained matter field can be calculated\cite{nphys,prl98}.
This is the most significant difference with the Berera's
formalism.

In this formalism a semiclassical expansion of the inflaton field
was proposed\cite{prd96}
\begin{displaymath}
\varphi(\vec x,t)=\phi_c (t)+ \phi(\vec x,t).
\end{displaymath}
Here the expectation value of the field operator $\varphi$, in some
unknown state $|E>$, is given by the classical field $\phi_c(t)$
and the expectation value for the quantum fluctuations $\phi(\vec x,t)$
is zero
\begin{displaymath}
<E|\varphi(\vec x,t)|E> = \phi_c(t); \qquad <E|\phi(\vec x,t)|E>=0.
\end{displaymath}
The field $\phi_c$ is responsible for the 
expansion of the universe,
while the quantum fluctuations $\phi(\vec x,t)$ takes into account
the local fluctuations of the matter field with respect to $\phi_c$.

\section{Formalism}

The Lagrangian that
describes the warm inflation scenario is
\begin{equation}
{\cal L}(\varphi,\varphi_{,\mu}) = - \sqrt{-g}\left[\frac{R}{16 \pi}
+\frac{1}{2} g^{\mu \nu} \varphi_{,\mu} \varphi_{,\nu}
+V(\varphi) \right]+{\cal L}_{int},
\end{equation}
where $R$ is the scalar curvature, $g^{\mu\nu}$ the metric
tensor and $g$ is the metric. The Lagrangian ${\cal L}_{int}$ 
takes into account
the interaction of the field $\varphi$ with other fields of the
thermal bath 
with mean temperature $<T_r> \  < T_{GUT} \sim 10^{15}$ GeV. 
This lower temperature condition implies that magnetic monopole
suppression works effectively.
As was recently showed, certain string inspired models 
solve all the cosmological puzzles
and have a motivation from string theory\cite{B1,BK}.
All particlelike matter which existed before inflation
would have been dispersed by inflation.

We consider the metric
\begin{equation}\label{1}
ds^2=-dt^2+ a^2(t) d\vec x^2.
\end{equation}
The eq. (\ref{1}) represents a flat
Friedmann - Robertson - Walker (FRW) metric
for a globally flat, isotropic
and homogeneous universe.
The Hubble parameter
$H[\varphi(\vec x,t)]$ can be written as an expansion
around $H_c(\phi_c)$
\begin{equation}
H[\varphi(\vec x,t)]=H_c(\phi_c)+ 
\sum_{n=1}^{\infty} \frac{1}{n!} \  H^{(n)}(\phi_c) \  \phi^n(\vec x,t),
\end{equation}
where $H^{(n)}(\phi_c)={d^n H(\varphi) \over d \varphi^n}|_{\phi_c}
\equiv H^{(n)}(\phi_c)$.
The equation of motion for the operator $\varphi$ is
\begin{equation}\label{as}
\ddot\varphi - \frac{1}{a^2}\nabla^2 \varphi + \left[3 H(\varphi)+
\tau(\varphi)\right] \  \dot\varphi + V'(\varphi)=0.
\end{equation}
The semiclassical 
Friedmann equation for a globally flat universe is given by
\begin{equation}\label{as1}
\left<E\left| H^2(\varphi)\right|E\right>=\frac{8 \pi}{3 M^2_p} 
\left<E\left| \rho_m+\rho_r \right|E\right>,
\end{equation}
where $M_p$ is the Planckian mass. Here $\rho_m $ and $\rho_r$ are
\begin{eqnarray}
\rho_m(\varphi) &=& \frac{\dot\varphi^2}{2} + \frac{1}{a^2}
\left(\vec\nabla \varphi \right)^2+ V(\varphi), \label{as2} \\
\rho_r(\varphi) &=& \frac{\tau(\varphi)}{8 H(\varphi)} \dot\varphi^2. \label{as3}
\end{eqnarray}
The scalar potential $V(\varphi)$ also can be written as an expansion
around $V(\phi_c)$
\begin{equation}
V(\varphi) = V(\phi_c) + 
\sum_{n=1}^{\infty} \frac{1}{n!} \  V^{(n)}(\phi_c) \  \phi^n(\vec x,t),
\end{equation}
where $V^{(n)}(\phi_c) \equiv \left.{d^n V(\varphi) 
\over d\varphi^n}\right|_{\phi_c}$.
Here
the prime denotes the derivative with respect to the field $\varphi$.
Furthermore $V'(\varphi)$ can be expanded as
\begin{equation}
V'(\varphi) = V'(\phi_c) + 
\sum_{n=2}^{\infty} \frac{1}{n!} \  V^{(n)}(\phi_c) \  \phi^{n-1}(\vec x,t).
\end{equation}
In the following I will consider $H(\varphi) \equiv H_c(\phi_c)$ and
$V(\varphi)$ and $V'(\varphi)$ as first ordered expansions in $\phi$.

\subsection{Dynamics of the classical field}

The classical equation of motion for the field $\phi_c$ is
\begin{equation}
\ddot\phi_c+ [3H_c(\phi_c)+ \tau_c(\phi_c)] \dot\phi_c+ V'(\phi_c)=0,
\end{equation}
where
$V'(\phi_c)\equiv {d V(\varphi)\over d \varphi}|_{\phi_c}$ 
and $\tau(\varphi) \equiv \tau_c(\phi_c)$.
Here, the laplacian term does not appears, since $\phi_c$ only depends
on time.
The evolution of $\phi_c$
and $H_c(\phi_c)$ are\cite{plb98}
\begin{eqnarray}
\dot\phi_c &=& - \frac{M^2_p}{4\pi}H'_c 
\left(1+ \frac{\tau_c}{3 H_c}
\right)^{-1}, \label{b}\\
\dot H_c & =& H'_c \dot\phi_c = - \frac{M^2_p}{4\pi} (H'_c )^2
\left(1+ \frac{\tau_c}{3 H_c}\right)^{-1},
\end{eqnarray}
where $H_c$ decreases with time due to the fact that $\dot H_c(t) <0$.
Furthermore, from eq. (\ref{as1}), the classical Friedmann equation is
\begin{equation}\label{al}
H^2_c(\phi_c)=\frac{4\pi}{3 M^2_p}\left[ \left(1+
\frac{\tau_c}{4 H_c}\right) \dot\phi^2_c+ 2 V(\phi_c)\right].
\end{equation}
Replacing (\ref{b}) in eq. (\ref{al}), one obtains the classical scalar
potential
\begin{equation}\label{a}
V(\phi_c) = \frac{3 M^2_p}{8\pi}\left[ H^2_c(\phi_c)-
\frac{M^2_p}{12\pi}\left(H'_c\right)^2 \left(1+\frac{\tau_c}{4 H_c}
\right)
\left(1+\frac{\tau_c}{3 H_c}\right)^{-2}\right].
\end{equation}
This expression relates the potential with the Hubble and friction
parameters.

On the other hand, the radiation energy density is
\begin{equation}\label{um}
\rho_r(\phi_c)  = \frac{\tau_c}{8 H_c}\left(\frac{M^2_p}{4\pi}\right)^2
(H'_c)^2 \left(1+ \frac{\tau_c}{3 H_c}\right)^{-2}.
\end{equation}
From eq. (\ref{um}) one obtains the mean temperature of the
thermal bath, in which there are particles that would be
dispersed during inflation 
\begin{equation}\label{16}
<T_r> \  \propto \  \left[\rho_r(\phi_c)\right]^{1/4}.
\end{equation}
The eq. (\ref{16}) gives the temperature of the background, but does
not describes the local fluctuations $\delta T_r/<T_r>$.

\subsection{Dynamics of the quantum perturbations}

I consider the quantum fluctuations of the matter field for
an universe with a FRW metric (\ref{1}). 
I will consider the Hubble and friction parameters as spatially
homogeneous [i.e., $H(\varphi) \equiv H_c(\phi_c)$ and $\tau(\varphi) \equiv
\tau_c(\phi_c)$].
The equation of motion for the quantum fluctuations
is
\begin{equation}\label{klein}
\ddot\phi - \frac{1}{a^2}\nabla^2\phi + [3 H_c + \tau_c ] \dot\phi 
+ V''(\phi_c) \phi =0,
\end{equation}
which can be simplified with the map
$\chi = e^{3/2 \int(H_c +\tau_c/3) dt} \  \phi $
\begin{equation}\label{fui}
\ddot\chi - a^{-2} \nabla^2 \chi - \frac{k^2_o}{a^2}\chi =0.
\end{equation}
Here, $k_o(t)$ is the wavenumber that separates the infrared
sector (with $k < k_o$)
and the short - wavelength sector (with $k > k_o$)
\begin{equation}
k^2_o(t)= a^2(t)\left[\frac{9}{4}\left(H_c(t)+\frac{\tau_c(t)}{3}\right)^2 
-V''_c(t)+\frac{3}{2}\left(\dot H_c(t)+\frac{\dot\tau_c(t)}{3}\right)\right].
\end{equation}
Here, the parameters $H_c(\phi_c)$ and $\tau_c(\phi_c)$
are evaluated in $t$, due to the fact that $\phi_c \equiv \phi_c(t)$.
The infrared sector describes the macrophysics of the universe (i.e.,
describes the universe in a cosmological scale), while
the short - wavelength sector takes into account the microphysics of it. 
The eq. (\ref{klein})  is a Klein - Gordon equation with a time
dependent parameter of mass $\mu(t) = {k_o \over a}$.

We can write the redefined field $\chi$ as a Fourier expansion
--- in terms of the modes $\xi_k e^{i \vec k. \vec x}$ --- in the
$k$ - space
\begin{equation}\label{a3}
\chi(\vec x,t) = \frac{1}{(2\pi)^{3/2}}
\int d^3k \  \left[a_k e^{i \vec k . \vec x}\xi_k(t)+ H.c\right],
\end{equation}
where $\xi_k(t)$ are the time dependent modes with wavenumber $k$.
The annihilation and creation operators $a_k$ and $a^{\dagger}_k$,
satisfy the following commutation relations
\begin{equation}
\left[a_k,a^{\dagger}_{k'}\right] = \delta^{(3)} (k-k'); \qquad
\left[a_k,a_{k'}\right] =\left[a^{\dagger}_k,a^{\dagger}_{k'}\right] =0.
\end{equation}
Furthermore, the operators $\chi$ and $\dot\chi$ satisfy 
\begin{equation}\label{a2}
\left[\chi(\vec x,t), \dot\chi(\vec x,t)\right]
= i \delta^{(3)}(\vec x - \vec {x'}).
\end{equation}
The interpretation for eq. (\ref{a2}) 
is just that $\chi$ and $\dot\chi$ are canonically
conjugate variable if $\vec x$ and $\vec x'$ are within the same
smeared point inside the light cone. Otherwise, $\chi$ and $\dot\chi$ can
be measured independently.

Replacing eq. (\ref{a3}) in eq. (\ref{a2}) one obtains the following
condition for the time dependent modes $\xi_k$
\begin{equation}
\xi_k\dot\xi^*_k - \dot\xi_k\xi^*_k = i.
\end{equation}
When the time dependent modes are real one obtains
\begin{equation}
\xi_k\dot\xi^*_k - \dot\xi_k\xi^*_k = 
\xi_k\dot\xi_k - \dot\xi_k\xi_k = 0,
\end{equation}
and the operators $\chi$ and $\dot\chi$ commute\cite{nphys,prl98}.
Thus, the field
$\chi$ becomes classical when all the modes $\xi_k(t)$ in eq. (\ref{a3})
are real. 
Note that $k_o$ depends on time and continuosly new time dependent
modes $\xi_k$ enters in the infrared sector ($k \ll k_o$)
from the short - wavelength sector ($k \gg k_o$).
We are interested in studying the sector of the 
universe where the redefined quantum
fluctuations $\chi$ become classical. Of course, this
condition imposes restrictions over the vacuum\cite{nphys,prl98} and
thus the asymptotic vacuum must be real in the infrared sector.
However, the modes $\xi_k$ --- for $k >  k_o(t)$ --- are
complex, and during inflation $\chi$ and $\dot\chi$
does not commute.

\subsection{The data COBE coarse - grained field}

In this work I define
a coarse - grained field which is obtained from the experimental
data\cite{4}
\begin{equation}\label{a4}
\chi_{Ccg}= \frac{1}{(2\pi)^{3/2}} \int d^3 k \  G(k,t)
\left[ a_k e^{i \vec k.\vec x} \xi_k(t)+ H.c.\right],
\end{equation}
where the suppression factor is
\begin{equation}\label{a5}
G(k,t)=\sqrt{\frac{1}{1+\left(\frac{k_o(t)}{k}\right)^N}},
\end{equation}
with $N=m-n$. Causality places
a strict constraint on the suppression index: $N\ge 4-n$.
A suppression factor like eq. (\ref{a5}) (with $n \sim 1$)
also has been found in a model
with cosmic strings plus cold or hot dark matter\cite{AS,be3}.

Furthermore, the  squared fluctuations for the COBE coarse - grained
field is\cite{liddle}
\begin{eqnarray} 
\left<E \left| \chi^2_{Ccg} \right| E\right> & =& \int^{\infty}_{0}
\frac{dk}{k} \  {\cal P}_{\chi_{Ccg}}(k) \label{nume} \\
& =& \frac{1}{2 \pi^2} \int^{k_o}_{0} dk \  k^2 \left|\xi_k(t)\right|^2
G^2(k,t),
\end{eqnarray}
where the power spectrum ${\cal P}_{\chi_{Ccg}}(k)$
when
the horizon exit is assumed as\cite{be3}
\begin{equation}\label{28}
{\cal P}_{\chi_{Ccg}}(k) = A(t_*) \  \left(\frac{k}{k_o(t_*)}\right)^n f(k).
\end{equation}
The power spectrum ${\cal P}_{\chi_{Ccg}}$ contains four parameters:
the amplitude $A$, the spectral index $n$, the
suppression wavenumber $k_o$, and the suppression index $m$.
Furthermore, $t_*$ denotes the time when the horizon entry, for which
$k_o(t_*) \simeq \pi H_o$ in comoving scale.
Replacing eq. (\ref{a4}) in (\ref{fui}), and
neglecting the terms with $-{1\over a^2}\nabla^2 \chi_{cg}$, since
they are very small with respect to the another terms, 
one obtains the quantum stochastic equation for the COBE
coarse - grained field
\begin{equation}\label{a6}
\ddot\chi_{Ccg}- \left(\frac{k_o(t)}{a(t)}\right)^2 \  \chi_{Ccg}=
\frac{N}{k_o(t)} \left[ \xi_1(\vec x,t)+\xi_2(\vec x,t)\right].
\end{equation}
Here the noises $\xi_1$ and $\xi_2$ are given by 
\begin{eqnarray}
\xi_1(\vec x,t)&=&
-\frac{\dot k_o k^N_o}{(2\pi)^{3/2}} \int
d^3 k \  k^{-N} \  G^3(k,t) 
\left[ a_k e^{i \vec k.\vec x} \dot\xi_k(t)+ H.c.\right], \\
\xi_2(\vec x,t)&=&
\frac{k^{N-1}_o}{4 (2\pi)^{3/2}} \int
d^3 k \  k^{-N} \  G^5(k,t) 
\left[\left(\frac{k_o}{k}\right)^N 
\left(3 \dot k^2_o - 2 k_o \ddot k_o\right) + 2 \dot k^2_o (1-N)
- 2 k_o \ddot k_o \right] 
\nonumber \\
&\times & \left[ a_k e^{i \vec k.\vec x} \xi_k(t)+ H.c.\right].
\end{eqnarray}
The function $G(k,t)$ determines the stochastic
properties and spectrum of the noises $\xi_1 $ and $\xi_2 $.

Replacing eq. (\ref{a3}) in eq. (\ref{fui}), one obtains the equation of 
motion for the modes $\xi_k(t)$
\begin{equation}\label{iuo}
\ddot\xi_k(t)+ \omega^2_k \  \xi_k(t)=0,
\end{equation}
where $\omega_k(t)=a^{-1}[k^2-k^2_o(t)]^{1/2}$ is the oscillation
frequency of the modes. 
In a de Sitter expansion\cite{nc} these frequencies are constant.
Most generally, the frequencies depends on time. Note that
for $k < k_o$ the squared frequencies $\omega^2_k$ are positives
and the solutions of (\ref{iuo}) are real, but it does not occur
for $k > k_o$ where the solutions of (\ref{iuo}) become complex.

\subsection{Classicality conditions for the COBE coarse - grained field}

As in previous works\cite{nphys,prl98,plb98,prd96,nc} we 
are interested in studying the classicality
conditions for the quantum stochastic equation (\ref{a6}).
Observe that all the modes $\xi_k(t)$ on the infrared sector
are real.
If we write the modes as a complex function with
components $u_k(t)$ and $v_k(t)$
\begin{equation}
\xi_k(t)= u_k(t)+ {\rm i} \  v_k(t),
\end{equation}
the condition for that the modes to be real is
\begin{equation}
\alpha_k(t)= \left| \frac{v_k(t)}{u_k(t)}\right| \ll 1.
\end{equation}
The condition for the COBE coarse - grained field $\chi_{Ccg}$ to be
classical during inflation becomes\cite{nphys,prl98}
\begin{equation}\label{cond}
\frac{1}{M(t)}\sum^{k\simeq  k_o(t)}_{k=0} \alpha_k(t) \ll 1,
\end{equation}
where $M(t)$ is the time dependent number of degrees of freedom in
the infrared (large - wavelength) sector. When all the modes
of the infrared sector are real 
(i.e., when $\alpha_{k\simeq k_o} \ll 1$). 

\subsection{Heisenberg representation for the COBE coarse - grained field}

During inflation the field $\chi_{Ccg}$ obeys 
the quantum stochastic equation
\begin{equation}
\ddot\chi_{Ccg}- \left(\frac{k_o(t)}{a(t)}\right)^2 \  \chi_{Ccg}=
\frac{N}{k_o(t)} \left[ \xi_{c1}(\vec x,t)+\xi_{c2}(\vec x,t)\right],
\end{equation}
which can be written as
\begin{equation}\label{a7}
\ddot\chi_{Ccg}- \left(\frac{k_o(t)}{a(t)}\right)^2 \  \chi_{Ccg}+
\xi_c(\vec x, t)=0,
\end{equation}
where 
$\xi_c(\vec x, t)=- {N\over k_o(t)} \left[ \xi_{c1}(\vec x,t)
+\xi_{c2}(\vec x,t)\right]$. The new noises $\xi_{c1}$ and $\xi_{c2}$
are
\begin{eqnarray}
\xi_{c1}(\vec x,t)&=& 
-\frac{\dot k_o k^N_o}{(2\pi)^{3/2}} \int
d^3 k \  k^{-N} \  G^3(k,t) 
\left[ a_k e^{i \vec k.\vec x} \dot\xi_k(t)+
a^{\dagger}_k e^{-i \vec k.\vec x} \dot\xi_k(t)\right], \\
\xi_{c2}(\vec x,t)&=&
\frac{k^{N-1}_o}{4 (2\pi)^{3/2}} \int
d^3 k \  k^{-N} \  G^5(k,t) 
\left[\left(\frac{k_o}{k}\right)^N 
\left(3 \dot k^2_o - 2 k_o \ddot k_o\right) + 2 \dot k^2_o (1-N)
- 2 k_o \ddot k_o \right] 
\nonumber \\
&\times & \left[ a_k e^{i \vec k.\vec x} \xi_k(t)+ 
a^{\dagger}_k e^{-i \vec k.\vec x} \xi_k(t)\right].
\end{eqnarray}
These noises become from the increasing number of degrees of freedom
of the infrared sector, as a consequence of the modes which
enters in the infrared sector from the short - wavelength sector. 
In general, $\xi_{c1}$ is a colored noise, while $\xi_{c2}$ gives non - local
dissipation. Under special circumstances (i.e., for $N \rightarrow \infty$),
the noise $\xi_{c1}$ is nearly delta correlated and describes a nearly
white and gaussian noise. Furthermore, in this case $\xi_{c2}$ generates 
local dissipation.

The effective Hamiltonian associated with eq. (\ref{a7}) is
\begin{equation}\label{a8}
H_{eff}(\chi_{Ccg},t)= \frac{1}{2} P^2_{Ccg}+
\frac{1}{2} \mu^2(t) \  \chi^2_{Ccg} + \xi_c \  \chi_{Ccg},
\end{equation}
where $P_{Ccg}\equiv \dot\chi_{Ccg}$ and $\mu^2(t)={k^2_o\over a^2}$. 
Observe that $\xi_c$
plays the role of an external classical stochastic
force in the effective Hamiltonian
(\ref{a8}). Thus, one can write the Sch\"odinger equation
\begin{equation}\label{a9}
{\rm i}\frac{\partial }{\partial t} \psi(\chi_{Ccg},t)=
-\frac{1}{2} \frac{\partial^2}{\partial \chi^2_{Ccg}}
\psi(\chi_{Ccg},t) + \left[\frac{1}{2} \  \mu^2(t) \chi^2_{Ccg}+
\xi_c \  \chi_{Ccg} \right] \psi(\chi_{Ccg},t).
\end{equation}
Here, $\psi(\chi_{Ccg},t)$ is the wave function that characterize
the
fluctuations of the matter field
in the observable universe.
Since the time dependence of $\mu(t)$, the
Hamiltonian is non - conservative, also in the case in which one
would neglects the stochastic force. 
The only case where
$\mu$ does not present a time dependence is a de Sitter expansion
of the universe. In this case $\mu$ is constant\cite{nc}
and eq. (\ref{a8}) represents a harmonic oscillator with a stochastic
external force $\xi_c$. 
In this case we have a forced linear harmonic
oscillator and the solution is a coherent state with the displacement
due to the action of the external force\cite{mijic}.

The effective Hamiltonian (\ref{a8}) represents an oscillator
that experiences both,
the squeezing, due to the time dependent frequency, and the external
force, due to the inflow of the short modes in the infrared sector.

The probability to find the universe with a given $\chi_{Ccg}$
in a given time $t$ is
\begin{equation}
P(\chi_{Ccg},t)=\psi(\chi_{Ccg},t) \psi^*(\chi_{Ccg},t),
\end{equation}
where the asterisk denotes the complex conjugate.
The problem with the Heisenberg's representation is that is very
complicated to find solutions for $\psi(\chi_{Ccg},t)$.
However, it gives a clear conceptual notion of the physical problem
under consideration.

\section{Examples}

\subsection{de Sitter expansion: supercooled scenario}

I consider a de Sitter expansion for which
$H_c(\phi_c) \equiv H_o =$ constant.
In this case $H'=0$ and thus, from eq. (\ref{um}) one obtains
a supercooled expansion for the universe\cite{nc} ($\rho_r =0$).
This means
that the dissipation parameter is zero ($\tau_c = 0$) during the
expansion. From eq. (\ref{a}), one obtains a constant potential
\begin{equation}
V_o = \frac{3 M^2_p}{8 \pi} \  H^2_o.
\end{equation}
For a massive inflaton with mass $m$ the square parameter $\mu^2(t) =
{k^2_o(t) \over a^2}$ (where $a \propto e^{H_o t}$ ) also becomes
time independent
\begin{equation}
\mu^2 = \frac{9}{4} H^2_o - m^2.
\end{equation}
The equation for the time dependent modes $\xi_k(t)$ is
\begin{equation} 
\ddot \xi_k(t) + \left[ k^2 e^{-2 H_o t} - \frac{9}{4} H^2_o + m^2 \right]
\xi_k(t) = 0. 
\end{equation}
The general solution for $\xi_k$ is
\begin{equation}
\xi_k(t) = A_1 {\cal H}^{(1)}_{\nu} \left[\frac{k}{H_o}e^{-H_o t}\right]
+ A_2 {\cal H}^{(2)}_{\nu} \left[\frac{k} {H_o}e^{-H_o t}\right],
\end{equation}
where $\nu =  \sqrt{{9\over 4} - {m^2 \over  H^2_o}} < 3/2$.
Here ${\cal H}^{(1)}_{\nu}$ and ${\cal H}^{(2)}_{\nu}$ are the first and second
species Hankel functions. For sufficiently large $t$, i.e., for 
$t \gg H^{-1}_o$ one obtains the
expression for ${\cal H}^{(1)}_{\nu}$ and ${\cal H}^{(2)}_{\nu}$
\begin{equation}
{\cal H}^{(1,2)}_{\nu} \left[\frac{k}{H_o} e^{-H_o t} \right]
\simeq \frac{1}{\Gamma(1+\nu)}
\left(\frac{k }{2 H_o} e^{-H_o t}\right)^{\nu} \pm
\frac{i}{\pi} \Gamma(\nu)
\left(\frac{k }{2 H_o} e^{-H_o t}\right)^{-\nu}.
\end{equation}
Chosing $A_1 =0$, and from the relation $\xi_k \dot\xi^*_k -
\dot \xi_k \xi^*_k = i$, one obtains $A_2
= {i \over 2} \sqrt{{\pi \Gamma(1+\nu)\over \nu \Gamma(\nu) H_o}}$,
and thus
\begin{equation}
\xi_k(t) \simeq \frac{i}{2} \sqrt{\frac{\pi}{\nu H_o \Gamma(\nu)
\Gamma(\nu+1)}} \left[\frac{k}{2 H_o} e^{-H_o t} \right]^{\nu}
+
\frac{1}{2} \sqrt{\frac{\Gamma(\nu+1)\Gamma(\nu)}{\nu H_o \pi}}
\left[\frac{k}{2 H_o} e^{-H_o t} \right]^{-\nu}  .
\end{equation}
For very large $t$, one obtains the asymptotic modes
\begin{equation}
\left.\xi_k(t)\right|_{t \gg 1} \simeq
\frac{1}{2} \sqrt{\frac{\Gamma(\nu) \Gamma(\nu+1)}{\nu H_o \pi}}
\left[\frac{k}{2 H_o} e^{-H_o t} \right]^{-\nu}  .
\end{equation}
Thus, the squared fluctuations on the observable scale,
for very large $t$, are
\begin{equation}\label{hg}
\left<E\left| \phi^2_{Ccg} \right|E\right> \simeq
\frac{2^{2\nu - 3}}{\pi^3} \left[\frac{\Gamma(\nu+1)
\Gamma(\nu)}{\nu}\right]
H^{2\nu-1}_o \  e^{2(\nu - 3/2) H_o t}
\int^{k_p}_{0} dk \  k^{2(1-\nu)} \  G^2(k),
\end{equation}
where $\phi_{Ccg} = e^{-3/2 H_o t} \chi_{Ccg}$ and $k_p$ is
de wavenumber for the Planckian wavelength.
Comparing (\ref{hg}) with (\ref{nume}) and (\ref{28}),
one obtains the following relation for the spectral index
\begin{equation}
n-1 = 2 (1- \nu).
\end{equation}
The standard choice $n=1$ was first advocated by Harrison\cite{ha} and
Zel'dovich\cite{Ze} on the ground that it is scale invariant
at the epoch of horizon entry.
The constraint
\begin{equation}
|n-1| < 0.3,
\end{equation}
gives the following values for the mass of the inflaton field
\begin{equation}
1.36 \  H_o < m < 1.46 \  H_o.
\end{equation}
Thus, the parameter $\nu$ only can take the values
\begin{equation}\label{57}
0.351 \  < \  \nu \  < \  0.65.
\end{equation}
The power spectrum for the COBE coarse - grained field $\phi_{Ccg}$,
is
\begin{equation}
{\cal P}_{\phi_{Ccg}}(k) \propto k^{2(1 - \nu)} \  G^2(k).
\end{equation}
Furthermore, the amplitude $A(t)$ decreases with time
during the inflation era
\begin{equation}
A(t) = 
\frac{2^{2\nu-3}}{\pi^3} \left[\frac{\Gamma(\nu+1) \Gamma(\nu)}{\nu}
\right] \  H^{2\nu-1}_o \  e^{2(\nu - 3/2)H_o t},
\end{equation}
for $m \neq 0$. After the horizon entry (i.e., for $t > t_*$) the 
amplitude $A(t_*)$ becomes
\begin{equation}
A(t_*) = \frac{2^{2\nu-3}}{\pi^3} 
\left[\frac{\Gamma(\nu+1) \Gamma(\nu)}{\nu}
\right] \  H^{2\nu-1}_o \  e^{2(\nu - 3/2)H_o t_*}.
\end{equation}
Due to $|\delta_k|^2 = {\cal P}_{\phi_{Ccg}}(k)$\cite{liddle}, the spectral
density is $|\delta_k|= k^{1-\nu} G(k)$. From the condition (\ref{57}), one
obtains a positive exponent for $k$ in the spectral density. To obtain
a negative exponent in the spectral density $|\delta_k|$, the mass of the
inflaton field must be very small
\begin{equation}
m < \frac{\sqrt{5}}{2} \  H_o .
\end{equation}

\subsection{Power - law inflation}

In this example I consider a power - law expansion for the universe.
In this case the scale factor and the Hubble parameter are respectively
$a(t) \propto \left(t/t_o\right)^p$, and $H_c(t) = p/t$.
The temporal evolution of the classical field is 
$e^{-\phi_c(t)/m} =(H_o/p) t$. Furthermore, the
temporal evolution for the radiation energy density is [see eq. (\ref{um})]
\begin{equation}
\rho_r(t) =
\left(\frac{M^2_p}{4\pi}\right)^2 \frac{p^2}{m^2} 
\left(1+ \frac{\tau_c t}{3p}\right)^{-2} t^{-2}.
\end{equation}
The potential is given by 
\begin{equation}
V_c[\phi_c(t)] = \frac{3 M^2_p}{8 \pi} t^{-2}
\left[ p^2 - \frac{M^2_p}{12 \pi}\frac{p^2}{m^2} 
\left(1+ \frac{\tau_c t}{3p}\right)^{-1}
\left(1+ \frac{\tau_c t}{4p}\right)\right].
\end{equation}
The equation of motion for the modes is
\begin{equation}\label{xi}
\ddot\xi_k(t) + \left[ \frac{t^{2p}_o k^2}{t^{2p}} - \mu^2(t)\right]
\xi_k(t) =0,
\end{equation}
where
\begin{equation}\label{muu}
\mu^2(t) = \frac{9}{4} \left(p t^{-1} + \frac{\tau_c}{3} \right)^2 -
V''_c(t) - \frac{3}{2} \left( p t^{-2} - \frac{\dot\tau_c}{3}\right),
\end{equation}
and
\begin{eqnarray}
V''_c(t) &=& \frac{3 M^2_p p^5}{32 H^2_o} t^{-2} \frac{1}{\left[3 H_o p^3+
\tau^2_c m^2 H^3_o t^3\right]} \nonumber \\
&\times & \left[ 4 H^3_o p^2 \tau^2_c t^3 (4\tau_c - H_o) +
H^3_o t^2 p^{-2} \left[m^2 \tau_c \tau''_c - 2 m^2(\tau'_c)^2+ 144 p \tau^2_c +
6 m \tau_c \tau'_c - 45 \tau^2_c\right] \right. \nonumber \\
& + & \left. H^3_o p^{-1} t \left[ 3 m^2 \tau''_c +
432 p \tau_c + 6 m \tau'_c - 141 \tau_c\right] +
2 H^3_o (216 p - 77)\right]. \label{v2}
\end{eqnarray}
Here the prime denotes the derivative with respect to $\phi_c$ (i.e.,
$\tau'_c\equiv {d\tau_c \over d \phi_c}(t)$, etc.).
For simplicity I will consider $\tau_c = p \gamma t^{-1}$, where
$\gamma $ is a dimensionless constant. In this case the general solution
for the eq. (\ref{xi}) is
\begin{equation}\label{beta}
\xi_k(t) = A_1 \sqrt{t/t_o} {\cal H}^{(1)}_{\nu}\left[
\frac{k (t/t_o)^{1-p}}{H_o (p-1)}\right] +
A_2  \sqrt{t/t_o} {\cal H}^{(2)}_{\nu}\left[
\frac{k (t/t_o)^{1-p}}{H_o (p-1)}\right],
\end{equation}
where $\nu = {1\over 2(p-1)} \sqrt{1 + 4 K^2}$ and
\begin{equation}\label{M}
K^2 = \left[ \frac{9}{4} p^2 \left(1+\frac{\gamma}{3}\right)^2 -
\frac{3}{2} p \left(1+ \frac{\gamma}{3}\right) +
\frac{3 p^2 M^2_p}{2\pi m^2} \left(1-
\frac{M^2_p}{48 \pi m^2} \left(1+\frac{\gamma}{4}\right)
\left(1+ \frac{\gamma}{3}\right)^{-2}\right)\right].
\end{equation}
For $A_1 = 0$, the modes (\ref{beta}) for ${k (t/t_o)^{1-p} \over
H_o (p-1)} \ll 1$ are
\begin{equation}
\xi_k(t) \simeq \frac{\sqrt{t/t_o}}{2\sqrt{\pi}} \Gamma(\nu) 
\left[\frac{k (t/t_o)^{1-p}}{2(p-1) H_o}\right]^{-\nu},
\end{equation}
where $\Gamma(\nu)$ is the gamma function with argument $\nu$.

The square fluctuations on the observable scale are
\begin{equation}
\left< E | \phi^2_{Ccg}|E\right> \simeq \frac{\Gamma^2(\nu) 2^{2\nu} 
(p-1)^{2\nu}}{8\pi^3 (p-1)} \left(t/t_o\right)^{2\nu(p-1)+1}
H^{2\nu}_o \int^{k_o}_{0} dk \  k^{2(1-\nu)} G^2(k,t),
\end{equation}
where $\phi_{Ccg} = t^{-3/2p(1+\gamma/3)} \chi_{Ccg}$, and
\begin{equation}
G^2(k,t) = \frac{1}{\left[1+ \frac{K t^{p-1}}{k t^p_o}\right]^N}.
\end{equation}
From the condition $n-1 = 2(1-\nu)$ one obtains
\begin{equation}\label{T}
n-1 = 2 \left[ 1 - \frac{1}{2(p-1)} \sqrt{1+ 4 K^2}\right],
\end{equation}
where $K^2$ is given by eq. (\ref{M}).
The constraint $|n-1| < 0.3$ in eq. (\ref{T}) gives the following conditions
\begin{eqnarray}
K^2 &<& \frac{\left[ 1.3 (p-1)\right]^2 -1}{4}, \\
K^2 &>& \frac{\left[ 0.702 (p-1)\right]^2 -1}{4}.
\end{eqnarray}
For example, taking $p=4$, with the scale $m = 1$, 
one obtains the condition
\begin{equation}
7.1065 \  < \  \gamma \  < \  7.1286,
\end{equation}
which implies that
\begin{equation}\label{gamma}
7.1065 \  H_c(t) < \  \tau_c \  < \  7.1286 \  H_c(t).
\end{equation}
When the horizon entry, the condition (\ref{gamma}) becomes
\begin{equation}
7.1065 \  H_c(t_*) < \  \tau_c(t_*) \  < \  7.1286 \  H_c(t_*).
\end{equation}

\section{Conclusions}

In this thermal scenario the rapid cooling
followed by rapid heating of the standard inflation
is replaced by a smoothened
dissipative mechanism. 
The classical field $\phi_c(t)$ generates the
expansion of the scale factor of the universe. 
Furthermore the fluctuations of the matter field, with respect to the
homogeneous field $\phi_c(t)$, are described by the
field $\phi(\vec x,t)$.
In the framework of a more realistic treatment, the
fluctuations on the sub - Hubble scale are described
by the COBE coarse - grained field $\chi_{Ccg}$, which is
defined by a suppression factor $G(k)$, which tends to zero
for $k \rightarrow 0$. This field describes the observable
universe after the horizon entry.
The quantum to classical transition of the COBE coarse - grained
field $\chi_{Ccg}$, is due to the complex to real transition of
the modes $\xi_k$, during inflation. 
The infrared sector takes into account
only the modes with wavelength much bigger than the size of the
horizon $k^{-1} > k^{-1}_o(t)$. 
In particular, during inflation
$\chi_{Ccg}(\vec x,t)$ does not commutes with
$\dot\chi_{Ccg}(\vec x',t)$ for $|\vec x - \vec x'| < k^{-1}_o$.
This is due to $\chi$ and $\dot\chi$ are canonically
conjugate variables if are ``measured'' inside a
causally conected region of the spacetime. 
Otherwise, $\chi$ and $\dot\chi$ can
be ``measured'' independently. The now observable universe 
is composed by causally desconeted domains during inflation.

During inflation, the number of degrees
of freedom in the infrared sector $M(t)$ is constantly increasing,
since short - wavelengths modes cross the horizon
from the short - wavelength sector. This effect is seen in the
second order stochastic equation as a noise $\xi_c$ and 
produces quantum decoherence\cite{nambu} and non - local
dissipative effects
in the infrared sector. The stochastic properties of $\xi_c$
are given by the suppression factor $G$. Furthermore, the noise
$\xi_k$ can be seen as a stochastic force in the framework of
the Heisenberg's representation for the redefined quantum fluctuations
$\chi_{Ccg}$. 
The effective Hamiltonian in the Heisenberg's representation
describes a damped oscillator with a variable $\mu(t)$.
This oscillator experiences both,
squeezing, due to the time dependent $\mu(t)$, and a stochastic external
force, due to the inflow of the short - wavelength
modes in the infrared sector.

Finally, two examples were considered.
In the example for a de Sitter expansion of the
universe, the radiation component for the energy density becomes zero.
In this case the parameter of mass $\mu={k_o \over a}$ does not
depends on time. Furthermore, 
constraint $|n-1|<0.3$ for the spectrum index gives restrictions
for the mass of the inflaton field
($1.35 \  H_o < m < 1.46 \  H_o$) and the parameter $\nu$
($0.351 \  < \  \nu \  < \  0.65$).
Here, the spectral density being given by $|\delta_k| = k^{1-\nu} G(k)$
and the amplitude $A(t)$ decreases with
time as $A(t) \sim e^{2(\nu - 3/2)H_o t}$ for $m \neq 0$. When
the horizon entry (for $t = t_*$), this amplitude becomes
freezen with value $A(t_*)$.
In the example for a power - law expansion for the universe, the
constraint $|n-1| < 0.3$ for the spectrum index gives restrictions
for the friction parameter, which is now time dependent.
For the particular case $\tau_c = p \gamma t^{-1}$ and
$p=4$, one obtains that $7.1065 \  H_c(t_*) \  < \tau_c(t_*) <
7.1286 \  H_c(t_*)$.  

\vskip .5cm
\centerline{\bf Acknowledgements}
I thank Adnan Bashir for your careful reading of the manuscript and
CIC of Universidad Michoacana for financial
support in the form of a research grant. Also, I thank the referee
for comments and suggestions.

\end{document}